\documentclass[aps,floats,showpacs,nofootinbib,floatfix]{revtex4}%
\usepackage{eurosym}
\usepackage{amsfonts}
\usepackage{amsmath}
\usepackage{graphicx}
\usepackage{amssymb}%
{\catcode`\|=\active \gdef\Braket#1{\left<\mathcode`\|"8000\let|\bravert
{#1}\right>}}
\def\bravert{\egroup\,\vrule\,\bgroup}

\begin{document}
\title{The Pion-Photon Transition Distribution Amplitudes in the Nambu-Jona Lasinio Model}
\author{A. Courtoy}
\email{Aurore.Courtoy@uv.es}
\author{S. Noguera}
\email{Santiago.Noguera@uv.es}
\affiliation{Departamento de Fisica Teorica and Instituto de F\'{\i}sica Corpuscular,
Universidad de Valencia-CSIC, E-46100 Burjassot (Valencia), Spain.}
\date{\today }

\begin{abstract}

\end{abstract}
\begin{abstract}
We define the pion-photon Transition Distribution Amplitudes (TDA) in a field
theoretic formalism from a covariant Bethe-Salpeter approach for the
determination of the bound state. We apply our formalism to the Nambu~-~Jona
Lasinio model, as a realistic theory of the pion. The obtained vector and
axial TDAs satisfy all features required by general considerations. In
particular, sum rules and polynomiality condition are explicitly verified. We
have numerically proved that the odd coefficients in the polynomiality
expansion of the vector TDA vanish in the chiral limit. The role of PCAC and
the presence of a pion pole are explicitly shown.

\end{abstract}

\maketitle

\section{Introduction}

Hard reactions provide important information for unveiling the structure of
hadrons. The large virtuality, $Q^{2}$, involved in the processes allows the
factorization of the hard (perturbative) and soft (non-perturbative)
contributions in their amplitudes. Therefore these reactions are receiving
great attention by the hadronic physics community. In the past, only total
cross sections of inclusive processes or longitudinal asymmetries, that have
simple parton model interpretations, were studied. The basic theoretical
ingredients to be understood are diagonal Parton Distribution Functions (PDF)
\cite{Jaffe:1996zw}, governing the deep inclusive processes. In the recent
years a new variety of processes, like the Deeply Virtual Compton Scattering,
has been considered. These processes are governed by the Generalized Parton
Distributions (GPD) \cite{Mueller:1998fv, Radyushkin:1996nd,Ji:1996ek,
Diehl:2003ny}. From a theoretical point of view, PDFs are related to diagonal
matrix elements of a bilocal operator from the initial hadron state to the
same final hadron state (the same particle with the same momentum). The GPDs
are related to the matrix elements of the same bilocal operator as in the
previous case, where the initial and final hadron are the same particle, but
have different momentum. The GPDs describe non-forward matrix elements of
light-cone operators and therefore measure the response of the internal
structure of the hadrons to the probes.

The generalization of parton distributions to the case where the initial and
final states correspond to different particles has recently been proposed in
\cite{Pire:2004ie, Lansberg:2006fv}. Such distributions are called Transition
Distribution Amplitudes (TDA) since they have been introduced through
hadron-photon transitions. In particular, the easiest case is to consider
pion-photon TDA, governing processes like $\pi^{+}\pi^{-}\rightarrow
\gamma^{\ast}\gamma$ or $\gamma^{\ast}\pi^{+}\rightarrow\gamma\pi^{+}$ in the
kinematical regime where the virtual photon is highly virtual but with small
momentum transfer.

The aim of this work is to calculate the pion-photon TDAs in a field theoretic
scheme treating the pion as a bound state in a fully covariant manner using
the Bethe-Salpeter equation. In this way we preserve all invariances of the
problem. In order to perform a numerical study we will use the Nambu~-~Jona
Lasinio (NJL) model to describe the pion structure. The NJL model is the most
realistic model for the pion based on a local quantum field theory built with
quarks. It respects the realizations of chiral symmetry and gives a good
description of the low energy physics of the pion \cite{Klevansky:1992qe}.

The NJL model is a non-renormalizable field theory and therefore a cut-off
procedure has to be defined. We have chosen the Pauli-Villars regularization
procedure because it respects all the symmetries of the problem. The NJL model
together with its regularization procedure is regarded as an effective theory
of QCD. Moreover, it has been used to tune coefficients of Chiral Perturbation
Theory \cite{Bijnens:1995ww}.

The NJL model has been used to describe the soft (non perturbative) part of
the deep processes, while for the hard part conventional perturbative QCD must
be used. It has been applied to the study of pion PDF \cite{Davidson:1994uv,
Davidson:2001cc, Ruiz Arriola:2002bp} and to the pion GPD
\cite{Theussl:2002xp}. In the chiral limit, its quark valence distribution is
as simple as $q(x)=\theta\left(  x\right)  ~\theta(1-x)$. Once evolution is
taken into account, a good agreement is reached between the calculated PDF and
the experimental one \cite{Davidson:1994uv}. A more elaborated study of pion
PDF is done in Ref. \cite{Noguera:2005cc} using non local lagrangians
\cite{Noguera:2005ej}, which confirms that the result obtained in the NJL
model for the PDF is a good approximation.

As GPDs, TDAs must satisfy sum rules. The vector and axial TDA are connected
to the vector and axial form factors, $F_{V}$ and $F_{A},$ appearing in the
$\pi^{+}\rightarrow\gamma e^{+}\nu$\ process.
In order to have a proper understanding of the axial TDA we will make the
difference between two contributions to the axial current coming from the
analysis of the amplitude for the pion radiative decay. The first one is
originated from the internal structure of the hadron, in our case a pion. This
contribution is the proper axial TDA. A second contribution is present and it
can be understood as a manifestation of PCAC. The axial current can be coupled
to a pion. This pion is a virtual one and its contribution will be present
independently of the external hadron.

The polynomiality condition is also satisfied. We observe that, in the
polynomial expansion of the moments of the vector TDA, the odd powers of $\xi$
are chirally suppressed and that they vanish in the chiral limit.

Previous studies of the axial and vector pion-photon TDA have been done using
different quark models \cite{Tiburzi:2005nj, Broniowski:2007fs}. Both works
parametrize the TDAs by means of Double Distributions.

This paper is organized as follows. In section \ref{SecIITDA} we establish the
connection between the TDA and the vector and axial pion form factors, $F_{V}$
and $F_{A}$. In section \ref{SecIIIFTTAD} we define our approach for the TDA
and we calculate them in the NJL model. In section \ref{SecIVSRP} we study the
sum rules and the polynomiality condition of the TDAs. In section \ref{SecVDC}
we discuss our results and we finally give our conclusions in section
\ref{SecVIC}.

\section{The pion-photon Transition Distribution Amplitudes}

\label{SecIITDA}

\begin{figure}
[ptb]
\begin{center}
\includegraphics[
height=2.1741cm,
width=4.9071cm
]%
{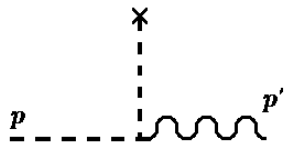}%
\caption{Pion pole contribution between the axial current (represented by a
cross) and the photon-external pion vertex associated to the last contribution
of equation (\ref{2.02}).}%
\label{Fig1}%
\end{center}
\end{figure}

The pion-photon TDAs are connected, through sum rules, to the vector and
axial-vector pion form factors, $F_{V}$ and $F_{A}$. Before giving a proper
definition of the TDAs let us recall the definition of these form factors.
They appear in the vector and axial vector hadronic currents contributing to
the decay amplitude of the process $\pi^{+}\rightarrow\gamma e^{+}\nu$. The
precise definitions of these currents are \cite{Moreno:1977kx, Bryman:1982et}:%
\begin{equation}
\left\langle \gamma\left(  p^{\prime}\right)  \right\vert \bar{q}\left(
0\right)  \gamma_{\mu}\tau^{-}q\left(  0\right)  \left\vert \pi\left(
p\right)  \right\rangle =-i\,e\,\varepsilon^{\nu}\,\epsilon_{\mu\nu\rho\sigma
}\,p^{\prime\rho}\,p^{\sigma}\,\frac{F_{V}\left(  t\right)  }{m_{\pi}}~,
\label{2.01}%
\end{equation}%
\begin{align}
\left\langle \gamma\left(  p^{\prime}\right)  \right\vert \bar{q}\left(
0\right)  \gamma_{\mu}\gamma_{5}\tau^{-}q\left(  0\right)  \left\vert
\pi\left(  p\right)  \right\rangle  &  =e\,\varepsilon^{\nu}\left(  p_{\mu
}^{\prime}\,p_{\nu}-g_{\mu\nu}\,p^{\prime}.p\right)  \frac{F_{A}\left(
t\right)  }{m_{\pi}}\nonumber\\
&  +e\,\varepsilon^{\nu}\left(  \left(  p^{\prime}-p\right)  _{\mu}\,p_{\nu
}\frac{2\sqrt{2}f_{\pi}}{m_{\pi}^{2}-t}-\sqrt{2}f_{\pi}\,g_{\mu\nu}\right)  ~,
\label{2.02}%
\end{align}
with $f_{\pi}=93%
\operatorname{MeV}%
,$ $\varepsilon^{0123}=1$ and $\tau^{-}=\left(  \tau_{1}-i\,\tau_{2}\right)
/2.$ All the structure of the decaying pion is included in the form factors
$F_{V}$ and $F_{A}$. We observe that the vector current only contains a
Lorentz structure associated with the $F_{V}$ form factor. The axial current
is composed of two terms. The first one, defining $F_{A}$, gives the structure
of the pion. The second one corresponds to the axial current for a point-like
pion. It has two different contributions. The first one corresponds to a
point-like coupling between the incoming pion, the outcoming photon and a
virtual pion which is coupled to the axial current. It is depicted in the
diagram of Fig. \ref{Fig1} and can be seen as a result of PCAC, because the
axial current must be coupled to the pion. It isolates the pion pole
contribution of the axial current in a model independent way. The second
contribution of this term, proportional to $f_{\pi}\,g_{\mu\nu}$, corresponds
to a pion-photon-axial current contact term. With these definitions, all the
structure of the pion remains in the form factor $F_{A}$.%

Let us go now to TDAs. For their definition we introduce the light-cone
coordinates $v^{\pm}=\left(  v^{0}\pm v^{3}\right)  /\sqrt{2}$ and the
transverse components $\vec{v}^{\bot}=\left(  v^{1},v^{2}\right)  $ for any
four-vector $v^{\mu}$. We define $P=\left(  p+p^{\prime}\right)  /2$\ and the
momentum transfer, $\Delta=p^{\prime}-p,$ therefore $P^{2}=m_{\pi}^{2}/2-t/4$
and $t=\Delta^{2}$. The skewness variable describes the loss of plus momentum
of the incident pion, i.e. $\xi=\left(  p-p^{\prime}\right)  ^{+}/2P^{+},$ and
its value ranges between $t/\left(  2m_{\pi}^{2}-t\right)  <\xi<1$. Actually
there is no symmetry relating the distributions for negative and positive
$\xi$ which could have constrained the values of the skewness variable to be
positive, like for GPDs. The vector and axial TDAs are the Fourier transform
of the matrix element of the bilocal currents, $\bar{q}\left(  -z/2\right)
\gamma_{\mu}\left[  \gamma_{5}\right]  q\left(  z/2\right)  $, separated by a
light-like distance. Then, they are directly related to the currents defined
in Eqs. (\ref{2.01}-\ref{2.02}) through the sum rules:%
\begin{equation}
\int_{-1}^{1}dx\int\frac{dz^{-}}{2\pi}e^{ixP^{+}z^{-}}\left.  \left\langle
\gamma\left(  p^{\prime}\right)  \right\vert \bar{q}\left(  -\frac{z}%
{2}\right)  \gamma_{\mu}\left[  \gamma_{5}\right]  \tau^{-}q\left(  \frac
{z}{2}\right)  \left\vert \pi\left(  p\right)  \right\rangle \right\vert
_{z^{+}=z^{\bot}=0}=\frac{1}{P^{+}}\left\langle \gamma\left(  p^{\prime
}\right)  \right\vert \bar{q}\left(  0\right)  \gamma_{\mu}\left[  \gamma
_{5}\right]  \tau^{-}q\left(  0\right)  \left\vert \pi\left(  p\right)
\right\rangle ~. \label{2.03}%
\end{equation}
With this connection we can introduce the leading twist decomposition of the
bilocal currents. For that we introduce the light-front vectors $\bar{p}^{\mu
}=P^{+}\left(  1,0,0,1\right)  /\sqrt{2}$ and $n^{\mu}=\left(
1,0,0,-1\right)  /\left(  \sqrt{2}P^{+}\right)  .$ The explicit expressions
for the pion and photon momenta in terms of their light-cone components are
given by Eqs. (\ref{plc}) and (\ref{pplc}).
Then we have%
\begin{equation}
\int\frac{dz^{-}}{2\pi}e^{ixP^{+}z^{-}}\left.  \left\langle \gamma\left(
p^{\prime}\right)  \right\vert \bar{q}\left(  -\frac{z}{2}\right)
\rlap{$/$}n\hspace*{-0.05cm}\tau^{-}q\left(  \frac{z}{2}\right)  \left\vert
\pi\left(  p\right)  \right\rangle \right\vert _{z^{+}=z^{\bot}=0}=\frac
{i}{P^{+}}\,e\,\varepsilon^{\nu}\,\epsilon_{\mu\nu\rho\sigma}\,n^{\mu
}\,P^{\rho}\,\Delta^{\sigma}\,\frac{V^{\pi^{+}}\left(  x,\xi,t\right)  }%
{\sqrt{2}f_{\pi}}~, \label{2.04}%
\end{equation}%
\begin{align}
\int\frac{dz^{-}}{2\pi}e^{ixP^{+}z^{-}}\left.  \left\langle \gamma\left(
p^{\prime}\right)  \right\vert \bar{q}\left(  -\frac{z}{2}\right)
\rlap{$/$}n\hspace*{-0.05cm}\gamma_{5}\tau^{-}q\left(  \frac{z}{2}\right)
\left\vert \pi\left(  p\right)  \right\rangle \right\vert _{z^{+}=z^{\bot}=0}
&  =\frac{1}{P^{+}}e\,\left(  \vec{\varepsilon}^{\bot}\cdot\vec{\Delta}^{\bot
}\right)  \frac{A^{\pi^{+}}\left(  x,\xi,t\right)  }{\sqrt{2}f_{\pi}%
}\nonumber\\
&  +\frac{1}{P^{+}}e\,\left(  \varepsilon\cdot\Delta\right)  \frac{2\sqrt
{2}f_{\pi}}{m_{\pi}^{2}-t}~\epsilon\left(  \xi\right)  ~\phi\left(
\frac{x+\xi}{2\xi}\right)  \quad, \label{2.05}%
\end{align}
where $\epsilon\left(  \xi\right)  $ is equal to $1$ for $\xi>0,$ and equal to
$-1$ for $\xi<0$. Here $V\left(  x,\xi,t\right)  $\ and $A\left(
x,\xi,t\right)  $ are respectively the vector and axial TDAs. They are defined
as dimensionless quantities. From the condition (\ref{2.03}) we observe that
they obey the following sum rules:%
\begin{equation}
\int_{-1}^{1}dx~V^{\pi^{+}}\left(  x,\xi,t\right)  =\frac{\sqrt{2}f_{\pi}%
}{m_{\pi}}F_{V}\left(  t\right)  ~, \label{2.06}%
\end{equation}%
\begin{equation}
\int_{-1}^{1}dx~A^{\pi^{+}}\left(  x,\xi,t\right)  =\frac{\sqrt{2}f_{\pi}%
}{m_{\pi}}F_{A}\left(  t\right)  ~, \label{2.07}%
\end{equation}
which were firstly introduced in Refs. \cite{Pire:2004ie,Lansberg:2006fv}.

In the second term of Eq. (\ref{2.05}), we have introduced the Pion
Distribution Amplitude (PDA) $\phi\left(  x\right)  .$ By definition the PDA
is%
\begin{equation}
\int\frac{dz^{-}}{2\pi}e^{i\left(  x-\frac{1}{2}\right)  p^{+}z^{-}}\left.
\left\langle 0\right\vert \bar{q}\left(  -\frac{z}{2}\right)
\rlap{$/$}n\hspace*{-0.05cm}\gamma_{5}\tau^{-}q\left(  \frac{z}{2}\right)
\left\vert \pi\left(  p\right)  \right\rangle \right\vert _{z^{+}=z^{\bot}%
=0}=\frac{1}{p^{+}}i\sqrt{2}f_{\pi}\phi\left(  x\right)  \quad, \label{2.08}%
\end{equation}
The PDA vanishes outside the region $x\in\left[  0,1\right]  $ and satisfies
the normalization condition%
\begin{equation}
\int_{0}^{1}dx~\phi\left(  x\right)  =1~. \label{2.09}%
\end{equation}%
\begin{figure}
[ptb]
\begin{center}
\includegraphics[
height=4.8607cm,
width=6.3494cm
]%
{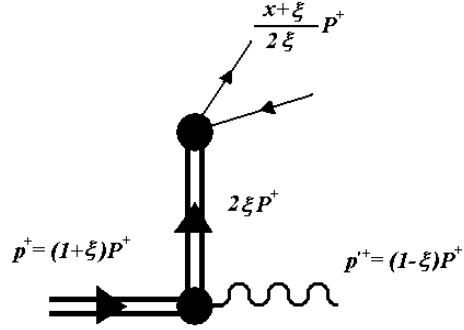}%
\caption{Pion pole contribution to the axial bilocal current corresponding to
the last term of equation (\ref{2.05}).}%
\label{Fig2}%
\end{center}
\end{figure}
This second term has been introduced in order to isolate the pion pole
contribution of the axial current in a model independent way, as we have done
in Eq. (\ref{2.02}) for the $\pi^{+}\rightarrow\gamma e^{+}\nu$ process.
Therefore, all the structure of the pion remains in the TDA $A\left(
x,\xi,t\right)  $. It can be seen as a result of PCAC, because the axial
current must be coupled to the pion. Therefore, this term is not a peculiarity
of the pion-photon TDAs. A similar pion term will be present in the Lorentz
decomposition in terms of distribution amplitudes of the axial current for any
pair of external particles. A pion exchange contribution has already been
analyzed in \cite{Mankiewicz:1998kg, Penttinen:1999th} for the axial
helicity-flip GPD and, in \cite{Tiburzi:2005nj}, a similar structure for the
axial current has been obtained using different arguments\footnote{In order to
make this connection it must be realized that there is a $\sqrt{2}$ \ between
our definition of $f_{\pi}$ and the one used in \cite{Tiburzi:2005nj}, and
that $\vec{\varepsilon}^{\bot}.\vec{\Delta}^{\bot}=\left(  1-\xi\right)
\left(  \varepsilon.\Delta\right)  $}. This term we have represented in Fig.
\ref{Fig2} is only non-vanishing in the ERBL region, i.e. the $x\in\left[
-\xi,\xi\right]  $ region. The kinematics of this region allow the emission or
absorption of a pion from the initial state, which is described through the
PDA. And it can be seen from Fig. \ref{Fig2} that positive values of $\xi$
corresponds to an outcoming virtual pion, whereas negative values of $\xi$
describe an incoming virtual pion. The latter is related to the matrix element
$\left\langle \pi\left(  p\right)  \right\vert \bar{q}\rlap{$/$}n\gamma
_{5}\tau^{-}q\left\vert 0\right\rangle ,$ instead of the one present in Eq.
(\ref{2.08}), what gives rise to the minus sign included in $\epsilon\left(
\xi\right)  .$

\section{A field theoretic approach to the pion-photon TDA.}

\label{SecIIIFTTAD}

In Ref. \cite{Theussl:2002xp} we have defined a method of calculation for the
pion GPD in a field theoretical scheme, treating the pion as a bound state of
quarks and antiquarks in a fully covariant manner using the Bethe-Salpeter
equation. We apply here the same method for evaluating the pion-photon TDAs.
This method has enormous advantages because it preserves all the physical
invariances of the problem. Therefore, any property as sum rules or
polynomiality is preserved.

\begin{figure}
[ptb]
\begin{center}
\includegraphics[
height=4.0931cm,
width=16.501cm
]%
{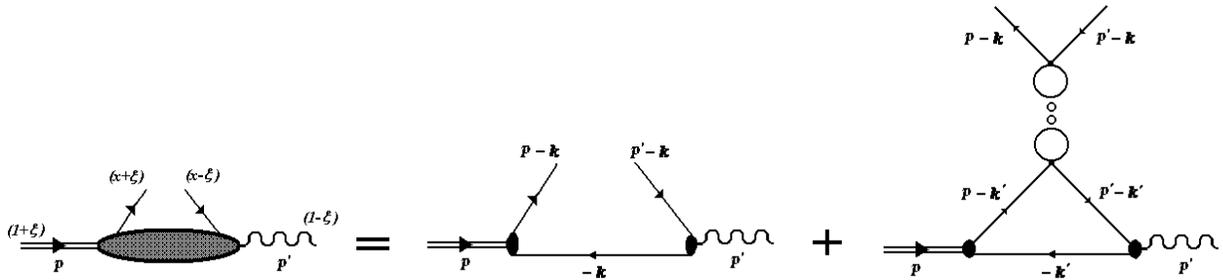}%
\caption{Diagrams contributing to the TDA. We have depicted diagrams in which
a quark $u$ is change into a quark $d$ by the bilocal current$.$ There are
similar diagrams in which the antiquark $\bar{d}$ is changed into a $\bar{u}$}%
\label{Fig3}%
\end{center}
\end{figure}

As usual, we consider that the process is dominated by the hand-bag diagram.
Each TDA has two related contributions, depending on which quark ($u$ or $d$)
of the pion is scattered off by the deep virtual photon. In Fig. \ref{Fig3} we
depicted the diagrams in which the photon scatters off the $u$-quark. We
observe that there are two kinds of contributing diagrams. In the first one
the $\bar{d}$-antiquark appears as the intermediate state, while in the second
the bi-local current couples to a quark-antiquark pair coupled in the pion
channel. The latter is present only for the axial current and includes the
pion pole contribution.%

The details of the method of calculation are given in Ref.
\cite{Theussl:2002xp}. In the present case we obtain, from the first kind of
diagram of Fig. \ref{Fig3}, the following contributions%
\begin{align}
&  \int\frac{dz^{-}}{2\pi}e^{ixP^{+}z^{-}}\left.  \left\langle \gamma\left(
p^{\prime},\varepsilon\right)  \left\vert \bar{q}\left(  -\frac{z}{2}\right)
\gamma^{\mu}\left[  \gamma_{5}\right]  \tau^{-}q\left(  \frac{z}{2}\right)
\right\vert \pi^{+}\left(  p\right)  \right\rangle \right\vert _{z^{+}%
=z^{\bot}=0}\nonumber\\
&  =-e\int\frac{d^{4}k}{\left(  2\pi\right)  ^{4}}\varepsilon_{\nu}\left\{
-\delta\left(  xP^{+}-\frac{1}{2}\left(  p^{\prime+}+p^{+}-2k^{+}\right)
\right)  \right. \nonumber\\
&  \mathbb{T}\mathrm{r}\left[  \frac{1}{2}\left(  \frac{1}{3}+\tau_{3}\right)
\,\gamma^{\nu}\,iS\left(  p^{\prime}-k\right)  \,\gamma^{\mu}\left[
\gamma_{5}\right]  \tau^{-}\,iS\left(  p-k\right)  \,\phi^{\pi^{+}}\left(
k,p\right)  \,iS\left(  -k\right)  \right] \nonumber\\
&  -\delta\left(  xP^{+}-\frac{1}{2}\left(  2k^{+}-p^{+}-p^{\prime+}\right)
\right) \nonumber\\
&  \left.  \mathbb{T}\mathrm{r}\left[  \frac{1}{2}\left(  \frac{1}{3}+\tau
_{3}\right)  \,\gamma^{\nu}\,iS\left(  k\right)  \,\phi^{\pi^{+}}\left(
k,p\right)  \,iS\left(  k-p\right)  \,\gamma^{\mu}\left[  \gamma_{5}\right]
\tau^{-}\,iS\left(  k-p^{\prime}\right)  \right]  \right\}  \quad,
\label{3.01}%
\end{align}
where $S\left(  p\right)  $ is the Feynman propagator of the quark and
$\phi^{\pi^{+}}\left(  k,p\right)  $ is the Bethe-Salpeter amplitude for the
pion. Here $\mathbb{T}\mathrm{r}()$ represents the trace over spinor, color
and flavor indices. The first contribution in Eq. (\ref{3.01}) is the one
depicted in the first diagram of Fig. \ref{Fig3}. The second contribution
corresponds to a similar diagram but changing quarks $u$ and $d$. In the NJL
model, $\phi^{\pi^{+}}\left(  k,p\right)  $ is as simple as%
\begin{equation}
\phi^{\pi^{+}}\left(  k,p\right)  =-g_{\pi qq}\,\gamma_{5}\,\sqrt{2}\tau^{+}~,
\label{3.02}%
\end{equation}
where $g_{\pi qq}$ is the pion-quark coupling constant defined in Eq.(
\ref{A.03}).

The vector TDA has contribution only from this first kind of diagrams. We can
express $V\left(  x,\xi,t\right)  $ as the sum of the active $u$-quark and the
active $\bar{d}$-quark distributions. The first contribution will be
proportional to the $d$'s charge, and the second contribution to the $u$'s
charge. Therefore, we can write%
\begin{equation}
V^{\pi^{+}}\left(  x,\xi,t\right)  =-\frac{1}{3}v_{u\rightarrow d}^{\pi^{+}%
}\left(  x,\xi,t\right)  +\frac{2}{3}v_{\bar{d}\rightarrow\bar{u}}^{\pi^{+}%
}\left(  x,\xi,t\right)  \quad. \label{3.03}%
\end{equation}
Isospin relates these two contributions, $v_{\bar{d}\rightarrow\bar{u}}%
^{\pi^{+}}\left(  x,\xi,t\right)  =v_{u\rightarrow d}^{\pi^{+}}\left(
-x,\xi,t\right)  .$ A direct calculation gives%
\begin{equation}
v_{u\rightarrow d}^{\pi^{+}}\left(  x,\xi,t\right)  =8\,N_{c}\,f_{\pi}\,g_{\pi
qq}\,m\,\tilde{I}_{3v}\left(  x,\xi,t\right)  \quad, \label{3.04}%
\end{equation}
with%
\begin{equation}
\tilde{I}_{3v}\left(  x,\xi,t\right)  =i\int\frac{d^{4}k}{\left(  2\pi\right)
^{4}}\,\delta\left(  x-1+\frac{k^{+}}{P^{+}}\right)  \frac{1}{\left(
k^{2}-m^{2}+i\epsilon\right)  \,\left[  \left(  p^{\prime}-k\right)
^{2}-m^{2}+i\epsilon\right]  \,\left[  \left(  p-k\right)  ^{2}-m^{2}%
+i\epsilon\right]  } \quad. \label{3.05}%
\end{equation}

In this integral, we first perform the integration over $k^{-}$. The pole
structure of the integrand fixes two non-vanishing contributions to
$v_{u\rightarrow d}^{\pi^{+}}\left(  x,\xi,t\right)  $, the first one in the
region $\xi<x<1$, corresponding to the quark contribution, and the second in
the region $-\xi<x<\xi,$ corresponding to a quark-antiquark contribution.
Given the relation (\ref{3.03}), the support of the entire vectorial TDA,
$V^{\pi^{+}}\left(  x,\xi,t\right)  $, is therefore $x\in\lbrack-1,1]$. The
analytical expression for (\ref{3.05}) is given by Eq. (\ref{A.05}). For the
$\pi^{0}$, the contributions of the first type of diagram (Fig. \ref{Fig3})
can be related to $v_{u\rightarrow d}^{\pi^{+}}\left(  x,\xi,t\right)  $%
\begin{align}
V_{u}^{\pi^{0}}  &  =\frac{Q_{u}}{\sqrt{2}}\left(  v_{u\rightarrow d}^{\pi
^{+}}\left(  x,\xi,t\right)  +v_{u\rightarrow d}^{\pi^{+}}\left(
-x,\xi,t\right)  \right)  \quad,\nonumber\\
V_{d}^{\pi^{0}}  &  =\frac{Q_{d}}{\sqrt{2}}\left(  v_{u\rightarrow d}^{\pi
^{+}}\left(  x,\xi,t\right)  +v_{u\rightarrow d}^{\pi^{+}}\left(
-x,\xi,t\right)  \right)  \quad. \label{vpi0}%
\end{align}

Turning our attention to the axial TDA, we find a new contribution arising
from the second diagram of Fig. \ref{Fig3}. This second contribution comes
from the re-scattering of a $q\bar{q}$ pair in the pion channel. Therefore%
\begin{align}
&  \int\frac{dz^{-}}{2\pi}e^{ixP^{+}z^{-}}\left.  \left\langle \gamma\left(
p^{\prime},\varepsilon\right)  \left\vert \bar{q}\left(  -\frac{z}{2}\right)
\gamma^{\mu}\gamma_{5}\tau^{-}q\left(  \frac{z}{2}\right)  \right\vert \pi
^{+}\left(  p\right)  \right\rangle \right\vert _{z^{+}=z^{\bot}=0}\nonumber\\
&  =(\text{\ref{3.01}})+\sum_{i}M^{i}\frac{2ig}{1-2g\Pi_{ps}\left(  t\right)
}N^{i}\quad, \label{3.06}%
\end{align}
where $i$ is an isospin index,%
\begin{align}
M^{i}  &  =-e\int\frac{d^{4}k^{\prime}}{\left(  2\pi\right)  ^{4}}%
\varepsilon_{\nu}\left\{  -\mathbb{T}\mathrm{r}\left[  \phi^{\pi^{+}}\left(
k^{\prime},p\right)  \,iS\left(  -k^{\prime}\right)  \frac{1}{2}\left(
\frac{1}{3}+\tau_{3}\right)  \,\gamma^{\nu}\,iS\left(  p^{\prime}-k^{\prime
}\right)  \,i\gamma_{5}\tau^{i}\,iS\left(  p-k^{\prime}\right)  \,\right]
\right. \nonumber\\
&  \left.  -\mathbb{T}\mathrm{r}\left[  \,\phi^{\pi^{+}}\left(  k^{\prime
},p\right)  \,iS\left(  k^{\prime}-p\right)  \,i\gamma_{5}\tau^{i}\,iS\left(
k^{\prime}-p^{\prime}\right)  \frac{1}{2}\left(  \frac{1}{3}+\tau_{3}\right)
\,\gamma^{\nu}\,iS\left(  k^{\prime}\right)  \right]  \right\}  ~,
\label{3.07}%
\end{align}%
\begin{equation}
N^{i}=-\int\frac{d^{4}k}{\left(  2\pi\right)  ^{4}}\delta\left(  xP^{+}%
-\frac{1}{2}\left(  p^{\prime+}+p^{+}-2k^{+}\right)  \right)  \mathbb{T}%
\mathrm{r}\left[  iS\left(  p^{\prime}-k\right)  \,\gamma^{\mu}\gamma_{5}%
\tau^{-}\,iS\left(  p-k\right)  \,i\gamma_{5}\tau^{i}\right]  \quad,
\end{equation}
and $\Pi_{ps}$ is the pseudoscalar polarization,%
\begin{equation}
\Pi_{ps}\left(  \Delta^{2}\right)  =-i\int\frac{d^{4}k}{\left(  2\pi\right)
^{4}}\left\{  \mathbb{T}\mathrm{r}\left[  i\gamma_{5}\,iS\left(  k\right)
\,i\gamma_{5}\,iS\left(  \Delta-k\right)  \right]  \right.  \quad.
\label{3.08}%
\end{equation}
We can now evaluate the axial current in a straightforward way. Nevertheless,
in order to extract the axial TDA we must subtract the pion pole contribution.
We need for that the pion amplitude which, in the NJL model, is
\begin{equation}
\phi\left(  x\right)  =\frac{m\,g_{\pi qq}\,N_{c}}{4\pi^{2}\,f_{\pi}}%
\,\sum_{i=1}^{2}c_{i}\log\frac{\left[  m^{2}-m_{\pi}^{2}x\left(  1-x\right)
\right]  }{\left[  m_{i}^{2}-m_{\pi}^{2}x\left(  1-x\right)  \right]  }%
\quad~~~~~~0\leq x\leq1~\quad, \label{3.09}%
\end{equation}
where $c_{i}$ and $m_{i}$ are defined in the Appendix. Now, after a long but
direct calculation, we obtain the expression for $A\left(  x,\xi,t\right)  .$
As the vector TDA, $A\left(  x,\xi,t\right)  $ can be expressed as a sum of
the contributions coming from the active $u$-quark and the active $\bar{d}%
$-quark. The first one will be proportional to the $d$'s charge and the second
to the $u$'s charge.%
\begin{equation}
A^{\pi^{+}}\left(  x,\xi,t\right)  =-\frac{1}{3}a_{u\rightarrow d}^{\pi^{+}%
}\left(  x,\xi,t\right)  +\frac{2}{3}a_{\bar{d}\rightarrow\bar{u}}^{\pi^{+}%
}\left(  x,\xi,t\right)  \quad. \label{3.10}%
\end{equation}
Isospin relates these two contributions, $a_{\bar{d}\rightarrow\bar{u}}%
^{\pi^{+}}\left(  x,\xi,t\right)  =-a_{u\rightarrow d}^{\pi^{+}}\left(
-x,\xi,t\right)  ,$ where the minus sign is originated in the change in
helicity produced by the $\gamma_{5}$ operator. The expression for
$a_{u\rightarrow d}^{\pi^{+}}\left(  x,\xi,t\right)  $ depends on the sign of
$\xi.$ In the $\left\vert \xi\right\vert <x<1$ region we have%
\begin{align}
a_{u\rightarrow d}^{\pi^{+}}\left(  x,\xi,t\right)   &  =-8\,N_{c}\,f_{\pi
}\,g_{\pi qq}\,m\left\{  \left(  1+\frac{m_{\pi}^{2}\left(  x-\xi\right)
+\left(  1-x\right)  t}{2\xi m_{\pi}^{2}-t\left(  1+\xi\right)  }\frac{1+\xi
}{1-\xi}\right)  \tilde{I}_{3v}\left(  x,\xi,t\right)  \right. \nonumber\\
&  +\frac{1}{2\xi m_{\pi}^{2}-t\left(  1+\xi\right)  }\left.  \frac{1+\xi
}{1-\xi}\frac{1}{16\pi^{2}}\sum_{i=1}^{2}c_{i}\left[  \log\frac{m^{2}\left(
m_{i}^{2}-\bar{z}\,m_{\pi}^{2}\right)  }{m_{i}^{2}\left(  m^{2}-\bar
{z}\,m_{\pi}^{2}\right)  }\right]  \right\}  \quad,
\end{align}
with the abbreviations $\bar{z}=\left(  1-x\right)  \left(  \xi+x\right)
/\left(  1+\xi\right)  ^{2}$. And in the $-\left\vert \xi\right\vert
<x<\left\vert \xi\right\vert $ region we have
\begin{align}
a_{u\rightarrow d}^{\pi^{+}}\left(  x,\xi,t\right)   &  =-8\,N_{c}\,f_{\pi
}\,g_{\pi qq}\,m\left\{  \left(  1+\frac{m_{\pi}^{2}\left(  x-\xi\right)
+\left(  1-x\right)  t}{2\xi m_{\pi}^{2}-t\left(  1+\xi\right)  }\frac{1+\xi
}{1-\xi}\right)  \tilde{I}_{3v}\left(  x,\xi,t\right)  \right. \nonumber\\
&  +\frac{\epsilon\left(  \xi\right)  }{1-\xi}\frac{1}{16\pi^{2}}\sum
_{i=1}^{2}c_{i}\left[  \frac{1+\xi}{2\xi m_{\pi}^{2}-t\left(  1+\xi\right)
}\log\frac{\left(  4\xi^{2}m^{2}-\bar{x}\,t\right)  \left(  m_{i}^{2}-\bar
{y}\,m_{\pi}^{2}\right)  }{\left(  4\xi^{2}m_{i}^{2}-\bar{x}\,t\right)
\left(  m^{2}-\bar{y}\,m_{\pi}^{2}\right)  }\right. \nonumber\\
&  +\left.  \frac{2}{t-m_{\pi}^{2}}\left.  \log\frac{\left(  4\xi^{2}%
m^{2}-\bar{x}\,t\right)  \left(  4\xi^{2}m_{i}^{2}-\bar{x}\,m_{\pi}%
^{2}\right)  }{\left(  4\xi^{2}m_{i}^{2}-\bar{x}\,t\right)  \left(  4\xi
^{2}m^{2}-\bar{x}\,m_{\pi}^{2}\right)  }\right]  \right\}  \quad, \label{3.11}%
\end{align}
where $\bar{x}=\left(  \xi^{2}-x^{2}\right)  $ and $\bar{y}=\bar{z}$ for
$\xi>0$ and $\bar{y}=0$ for $\xi<0$.

The expressions for both the vector and axial TDAs in the chiral limit, i.e.
$m_{\pi}=0$, are well defined. In particular, for $t=0$, we find the following
simple expression for $v_{u\rightarrow d}^{\pi^{+}}(x,\xi,t)$
\begin{equation}
v_{u\rightarrow d}^{\pi^{+}}(x,\xi,0)=\sqrt{2}\,f_{\pi}\,6\,F_{V}^{\pi^{+}%
\chi}(0)\,\left[  \theta(\xi^{2}-x^{2})\,\,\frac{x+\left\vert \xi\right\vert
}{2\left\vert \xi\right\vert (1+\left\vert \xi\right\vert )}\,+\,\theta
(x-\left\vert \xi\right\vert )\theta(1-x)\frac{1-x}{1-\xi^{2}}\right]  \quad,
\label{chiv}%
\end{equation}
where $F_{V}^{\pi^{+}\chi}(0)$ is the chiral limit of the vector pion form
factor at zero momentum transfer $F_{V}^{\pi^{+}\chi}(0)=\lim_{m_{\pi
}\rightarrow0}F_{V}^{\pi^{+}}(0)/m_{\pi}=0.17%
\operatorname{GeV}%
^{-1}$.

A similar expression is obtained for $a_{u\rightarrow d}^{\pi^{+}}(x,\xi,t)$
in the chiral limit for $t=0$
\begin{align}
a_{u\rightarrow d}^{\pi^{+}}(x,\xi,0)  &  =-\sqrt{2}\,f_{\pi}\,6\,F_{A}%
^{\pi^{+}\chi}(0)\,\left[  \theta(\xi^{2}-x^{2})\,\epsilon\left(  \xi\right)
\,\frac{\left(  \xi-x\right)  }{4\xi^{2}(1+\left\vert \xi\right\vert )}\left(
x+\xi+\left(  x-\xi\right)  \frac{\left\vert \xi\right\vert -\xi
}{(1+\left\vert \xi\right\vert )}\right)  \right.  \,\,\nonumber\\
&  +\,\left.  \theta(x-\left\vert \xi\right\vert )\theta(1-x)\frac
{(1-x)(x-\xi)}{(1-\xi^{2})(1-\xi)}\right]  \quad, \label{chia}%
\end{align}
with $F_{A}^{\pi^{+}\chi}(0)$ the axial form factor at $t=0$ in the chiral
limit $F_{A}^{\pi^{+}\chi}(0)=\lim_{m_{\pi}\rightarrow0}F_{A}^{\pi^{+}%
}(0)/m_{\pi}=F_{V}^{\pi^{+}\chi}(0)$.

\section{Sum rules and polynomiality}

\label{SecIVSRP}The vector TDA of the $\pi^{+}$ must obey the sum rule given
in Eq.~(\ref{2.06}) with the expression of the vector pion form factor in the
NJL model given by Eq.~(\ref{fv}). We have numerically recovered the sum rule
for different $t$ values. In particular we obtain the value $F_{V}^{\pi^{+}%
}(0)=0.0242$ for the vector form factor at $t=0$, which is in agreement with
the experimental value $F_{V}(0)=0.017\pm0.008$ given in \cite{Yao:2006px}.

The $\pi^{0}$ distribution must satisfy the following sum rule
\cite{Pire:2004ie}
\begin{equation}
\int_{-1}^{1}dx\,\left(  Q_{u}\,V_{u}^{\pi^{0}}(x,\xi,t)-Q_{d}\,V_{d}^{\pi
^{0}}(x,\xi,t)\right)  =\sqrt{2}\,f_{\pi}\,F_{\pi\gamma^{\ast}\gamma}%
(t)\qquad. \label{4.01}%
\end{equation}
A theoretical prediction of the $\pi^{0}$ form factor value is given in
\cite{Brodsky:1981rp}. In particular, at $t=0$, the value $F_{\pi\gamma^{\ast
}\gamma}(0)=0.272$ GeV$^{-1}$ is found. The neutral pion form factor is
directly related to the $\pi^{+}$ vector form factor so that the sum rule is
satisfied. We obtain the value $F_{\pi\gamma^{\ast}\gamma}(0)=0.244~$%
GeV$^{-1}$. In \cite{Gronberg:1997fj}, a dipole parametrization based on
experimental data is proposed for the $t$-dependence of $F_{\pi\gamma^{\ast
}\gamma}\left(  t\right)  $, obtaining for the dipole mass $\Lambda=0.77$ GeV.
We have found that, for small values of $t,$ the NJL neutral pion form factor
can be parametrized in a dipole form with $\Lambda=0.81$ GeV.

The axial TDA obeys the sum rules given by Eq.~(\ref{2.07}) with the axial
form factor given in the NJL model by Eq.~(\ref{faalltwists}). This sum rule
is satisfied for different $t$ values. The numerical results also coincide. In
particular we obtain $F_{A}^{\pi^{+}}(0)=0.0239$ for the axial form factor at
$t=0$, which is about twice the value $F_{A}^{\pi^{+}}(0)=0.0115\pm0.0005$
given by the PDG \cite{Yao:2006px}.

We expect TDAs to obey the polynomiality condition. However no time reversal
invariance enforces the polynomials to be even in the $\Delta$-momenta like
for GPDs. That means that the polynomials should be \textquotedblleft
complete\textquotedblright, i.e. they should include all powers in $\xi$,
\begin{equation}
\int_{-1}^{1}dx\,x^{n-1}\,V(x,\xi,t)=\sum_{i=0}^{n-1}\,C_{n,i}(t)\,\xi
^{i}\quad. \label{oddeven}%
\end{equation}

\begin{table}[tb]
\centering{\small {
\begin{tabular}
[c]{|c|c|c|c|c|}\hline
$n$ & 1 & 2 & 3 & 4\\\hline
& {\scriptsize {$C_{1,0}\left(  t\right)  $}} & {\scriptsize {$%
\begin{array}
[c]{rr}%
C_{2,0}\left(  t\right)  & C_{2,1}\left(  t\right)
\end{array}
$ }} & {\scriptsize { $%
\begin{array}
[c]{rrr}%
C_{3,0}\left(  t\right)  & C_{3,1}\left(  t\right)  & C_{3,2}\left(  t\right)
\end{array}
$}} & {\scriptsize {$%
\begin{array}
[c]{rrrr}%
C_{4,0}\left(  t\right)  & C_{4,1}\left(  t\right)  & C_{4,2}\left(  t\right)
& C_{4,3}\left(  t\right)
\end{array}
$}}\\\hline\hline
{\footnotesize {$%
\begin{array}
[c]{c}%
m_{\pi}=0
\end{array}
$}} &  &  &  & \\\hline
{\footnotesize {$%
\begin{array}
[c]{l}%
t=0\\
t=-0.5\\
t=-1.0
\end{array}
$}} & $%
\begin{array}
[c]{l}%
22.6\\
13.7\\
\multicolumn{1}{r}{10.4}%
\end{array}
$ & $%
\begin{array}
[c]{ll}%
-22.6 & \hspace*{0.1cm}0.0\\
-16.3 & \hspace*{0.1cm}0.0\\
\multicolumn{1}{r}{-13.4} & \multicolumn{1}{r}{0.0}%
\end{array}
$ & $%
\begin{array}
[c]{lll}%
3.77 & \hspace*{0.3cm}0.0 & \hspace*{0.3cm}3.77\\
\multicolumn{1}{r}{3.00} & \hspace*{0.3cm}0.0 & \multicolumn{1}{r}{2.43}\\
\multicolumn{1}{r}{2.57} & \multicolumn{1}{r}{0.0} & \multicolumn{1}{r}{1.90}%
\end{array}
$ & $%
\begin{array}
[c]{llll}%
-6.7 & \hspace*{0.3cm}0.0 & \hspace*{0.3cm}-6.7 & \hspace*{0.3cm}0.0\\
-5.7 & \hspace*{0.3cm}0.0 & \hspace*{0.3cm}-4.9 & \hspace*{0.3cm}0.0\\
-5.0 & \multicolumn{1}{r}{0.0} & \multicolumn{1}{r}{-4.0} &
\multicolumn{1}{r}{0.0}%
\end{array}
$\\\hline\hline
{\footnotesize {$%
\begin{array}
[c]{c}%
m_{\pi}=140
\end{array}
$}} &  &  &  & \\\hline
{\footnotesize {$%
\begin{array}
[c]{l}%
t=0\\
t=-0.5\\
t=-1.0
\end{array}
$}} & $%
\begin{array}
[c]{l}%
22.7\\
\multicolumn{1}{r}{13.7}\\
\multicolumn{1}{r}{10.3}%
\end{array}
$ & $%
\begin{array}
[c]{ll}%
-22.9 & \hspace*{0.3cm}0.44\\
\multicolumn{1}{r}{-16.4} & \multicolumn{1}{r}{0.25}\\
\multicolumn{1}{r}{-13.4} & \multicolumn{1}{r}{0.19}%
\end{array}
$ & $%
\begin{array}
[c]{lll}%
3.80 & \hspace*{0.3cm}-0.10 & \hspace*{0.3cm}3.67\\
\multicolumn{1}{r}{3.00} & \multicolumn{1}{r}{-0.06} &
\multicolumn{1}{r}{2.43}\\
\multicolumn{1}{r}{2.57} & \multicolumn{1}{r}{-0.05} &
\multicolumn{1}{r}{1.87}%
\end{array}
$ & $%
\begin{array}
[c]{llll}%
-6.8 & \hspace*{0.2cm}0.19 & \hspace*{0.3cm}-6.8 & \hspace*{0.2cm}0.12\\
-5.7 & \hspace*{0.2cm}0.13 & \hspace*{0.3cm}-4.9 & \hspace*{0.2cm}0.08\\
-5.0 & \hspace*{0.2cm}0.11 & \hspace*{0.3cm}-4.0 & \multicolumn{1}{r}{0.06}%
\end{array}
$\\\hline
\end{tabular}
} }\caption{Coefficients of the polynomial expansion for the vector TDA. The
pion mass is expressed in MeV and $t$ is expressed in GeV$^{2}$. Notice that
the coefficients have to be multiplied by $10^{-3}$. }%
\label{tablepolyv}%
\end{table}However, in the chiral limit and for $t=0$, we have analytically
found that the odd powers in $\xi$ go to zero for the polynomial expansion of
the vector TDA. A study of the polynomiality in the limit given in
Eq.~(\ref{chiv}) leads to the following analytical expression for the
coefficients $C_{n,2i}^{\chi}(t)$
\begin{equation}
C_{n,2i}^{\chi}(0)=\sqrt{2}\,f_{\pi}\,\,2\,F_{V}^{\pi^{+}\chi}(0)\,\left(
-1+2(-1)^{n-1}\right)  \,\left(  \frac{1}{n}-\frac{1}{n+1}\right)  \quad.
\label{4.03}%
\end{equation}
Notice they do not depend on $i$. We have numerically tested the polynomiality
and obtained it. We observe that the coefficients for the odd powers in $\xi$
are of one order of magnitude smaller than those for the even powers in $\xi$.
In particular we have numerically proved that, in the chiral limit, the
polynomials only contain even powers in $\xi$ for any value of $t,$%
\begin{equation}
\int_{-1}^{1}dx\,x^{n-1}\,[\lim_{m_{\pi}\rightarrow0}V(x,\xi,t)]=\sum
_{i=0}^{[\frac{n-1}{2}]}\,C_{n,2i}^{\chi}(t)\,\xi^{2i}\quad. \label{even}%
\end{equation}
The coefficients in the chiral limit, Eq.~(\ref{even}), as well as the
coefficients for $m_{\pi}=140$ MeV, Eq.~(\ref{oddeven}), we numerically
obtained are given in Table~\ref{tablepolyv}.

The $\xi$-dependence of the moments of the axial TDA $A(x,\xi,t)$ also has a
polynomial form. Those polynomials contain all the powers in $\xi$, i.e. even
and odd,
\begin{equation}
\int_{-1}^{1}dx\,x^{n-1}\,A(x,\xi,t)=\sum_{i=0}^{n-1}\,C_{n,i}^{\prime
}(t)\,\xi^{i}\quad.
\end{equation}
An analytic study of the polynomiality in the limit given in Eq.~(\ref{chia})
confirms that all the powers in $\xi$ have to be present. The analytic values
for the coefficients in this specific limit are
\begin{align}
C_{n,2i}^{\prime\chi}(0)  &  = \sqrt{2}\,f_{\pi}\,\,2\,F_{A}^{\pi^{+}\chi
}(0)\,\left(  1+2(-1)^{n-1} \right)  \,\frac{n-2i}{n(n+1)(n+2)} \quad
,\nonumber\\
C_{n,2i+1}^{\prime\chi}(0)  &  =\sqrt{2}\,f_{\pi}\,\,2\,F_{A}^{\pi^{+}\chi
}(0)\,\left(  1+2(-1)^{n-1} \right)  \,\frac{-2(i+1)}{n(n+1)(n+2)}\quad,
\label{coefaxial}%
\end{align}
which is in agreement with those numerically obtained. The coefficients of the
polynomial expansions in the chiral limit are very close to those obtained for
the physical values of the pion mass. In Table~\ref{tablepolya}, the
coefficients $C_{n,i}^{\prime}(t)$ are therefore shown only for $m_{\pi}=140%
\operatorname{MeV}%
$. The polynomiality property of the term containing the pion DA can also be
studied. The $t$-dependence only comes from the pion pole. We can therefore
write a general relation about the polynomiality property of the whole axial
bilocal matrix element
\begin{equation}
\int_{-1}^{1}dx\,x^{n-1}\,\int\frac{dz^{-}}{2\pi}\,e^{ixp^{+}z^{-}}%
\,\langle\gamma(P^{\prime})|\bar{q}\left(  -\frac{z}{2}\right)  \,\gamma_{\mu
}n^{\mu}\gamma_{5}\,\tau^{-}q\left(  \frac{z}{2}\right)  |\pi(P)\rangle
\,\,|_{z^{+}=z^{\perp}=0}=\frac{1}{p^{+}}\,e\,(\varepsilon.\Delta)\,\frac
{1}{\sqrt{2}f_{\pi}}\,\sum_{i=0}^{n}\,C_{n,i}^{^{\prime\prime}}(t)\,\xi
^{i}\quad.
\end{equation}

\begin{table}[tb]
\centering
\begin{tabular}
[c]{|c|c|c|c|}\hline
$n$ & 1 & 2 & 3\\\hline
& $C^{\prime}_{1,0}\left(  t\right)  $ & $%
\begin{array}
[c]{rr}%
C^{\prime}_{2,0}\left(  t\right)  & C^{\prime}_{2,1}\left(  t\right)
\end{array}
$ & $%
\begin{array}
[c]{rrr}%
C^{\prime}_{3,0}\left(  t\right)  & C^{\prime}_{3,1}\left(  t\right)  &
C^{\prime}_{3,2}\left(  t\right)
\end{array}
$\\\hline\hline
$%
\begin{array}
[c]{c}%
m_{\pi}=140\,\,\,\,\,
\end{array}
$ &  &  & \\\hline
$%
\begin{array}
[c]{l}%
t=0\\
t=-0.5\\
t=-1.0
\end{array}
$ & $%
\begin{array}
[c]{l}%
22.4\\
\multicolumn{1}{r}{16.1}\\
\multicolumn{1}{r}{13.2}%
\end{array}
$ & $%
\begin{array}
[c]{ll}%
-3.77 & \hspace*{.2cm} 4.00\\
\multicolumn{1}{r}{-2.97} & \multicolumn{1}{r}{2.57}\\
\multicolumn{1}{r}{-2.53} & \multicolumn{1}{r}{1.97}%
\end{array}
$ & $%
\begin{array}
[c]{lll}%
6.8 & \hspace*{.2cm} -4.9 & \hspace*{.2cm} 2.3\\
\multicolumn{1}{r}{5.7} & \multicolumn{1}{r}{-3.5} & \multicolumn{1}{r}{1.7}\\
\multicolumn{1}{r}{5.0} & \multicolumn{1}{r}{-2.8} & \multicolumn{1}{r}{1.3}%
\end{array}
$\\\hline
\end{tabular}
\caption{Coefficients of the polynomial expansion for the axial TDA. The pion
mass is expressed in MeV and $t$ is expressed in GeV$^{2}$. Notice that the
coefficients have to be multiplied by $10^{-3}$. }%
\label{tablepolya}%
\end{table}

\section{Discussion.}

\label{SecVDC}

\begin{figure}[tb]
\centering
\includegraphics [height=5.5cm]{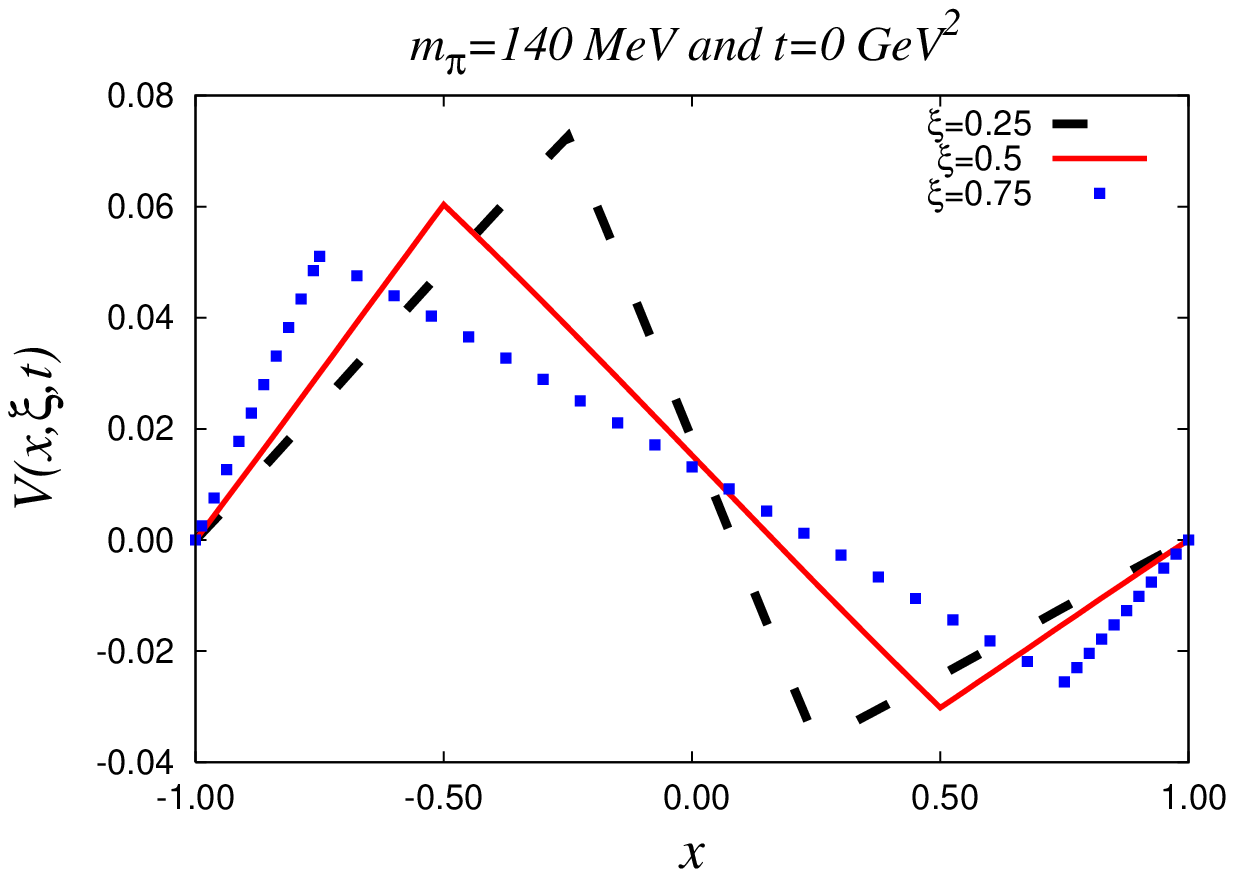}
\includegraphics [height=5.5cm]{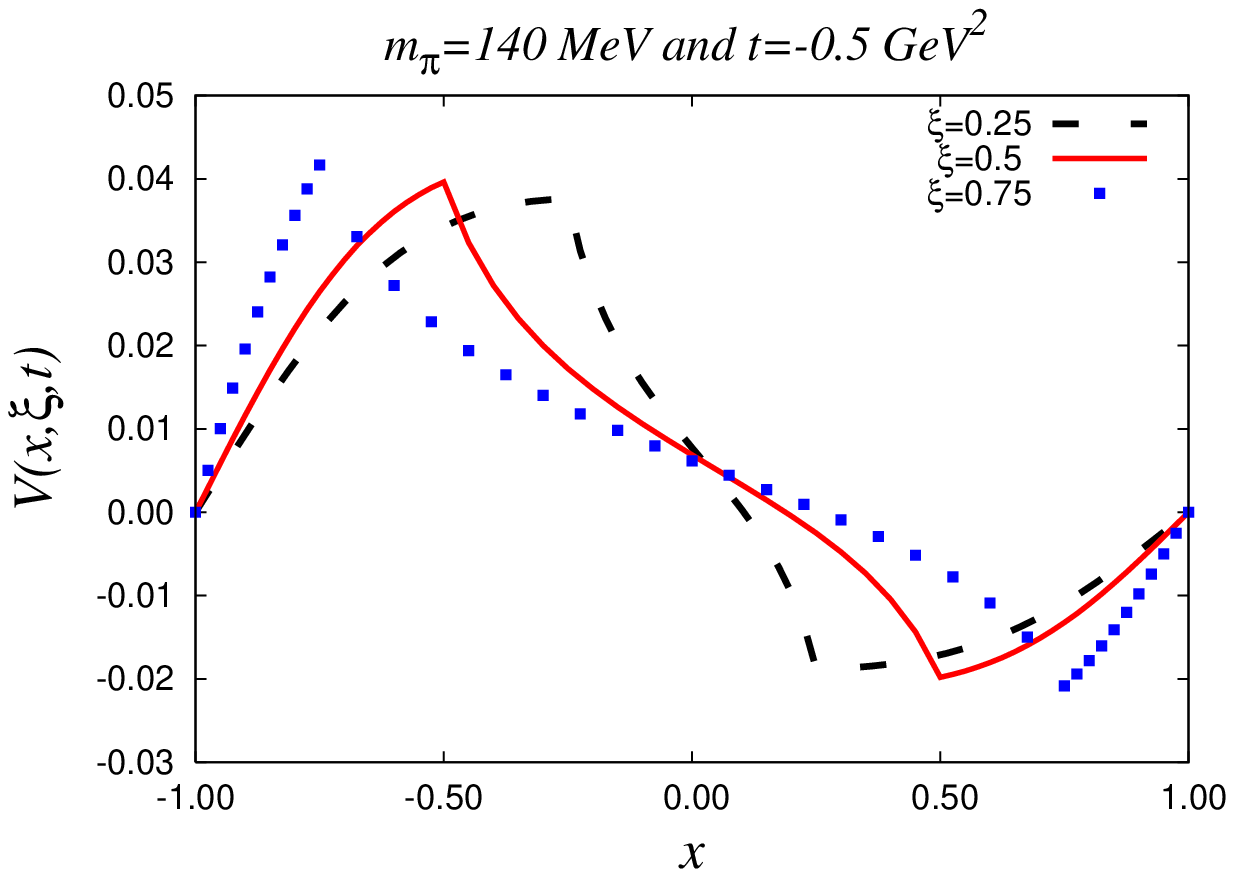}\caption{ The vector TDA in the case
$m_{\pi}=140$ MeV and, respectively, $t=0$ and $t=-0.5$ GeV$^{2}$.}%
\label{TDAmpi140}%
\end{figure}

\begin{figure}[tbtb]
\centering
\includegraphics [height=7cm]{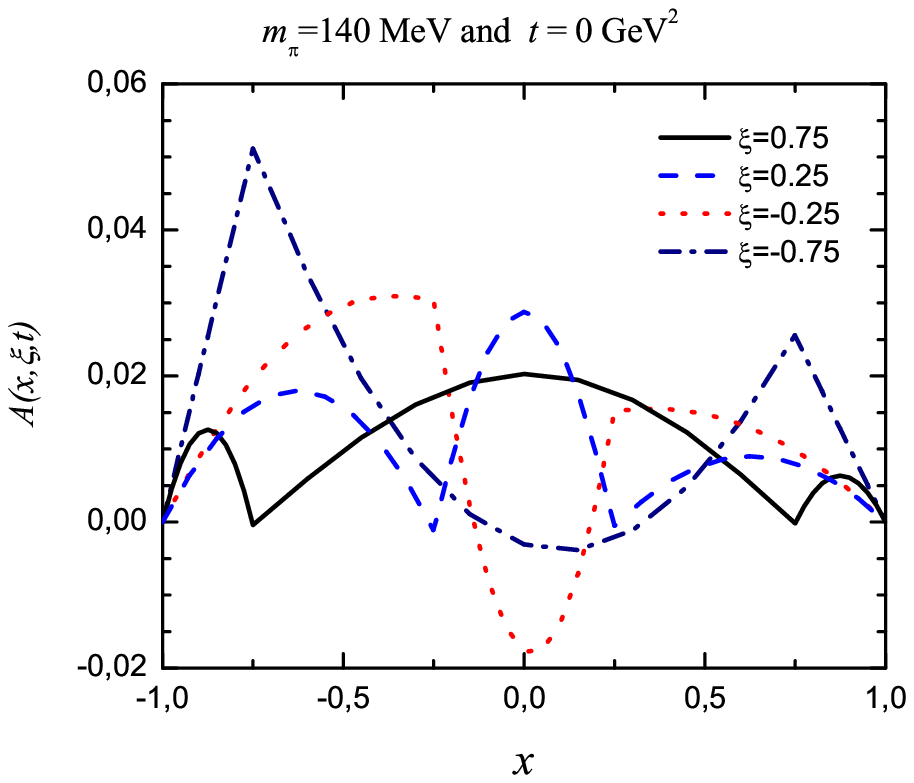}
\includegraphics [height=7cm]{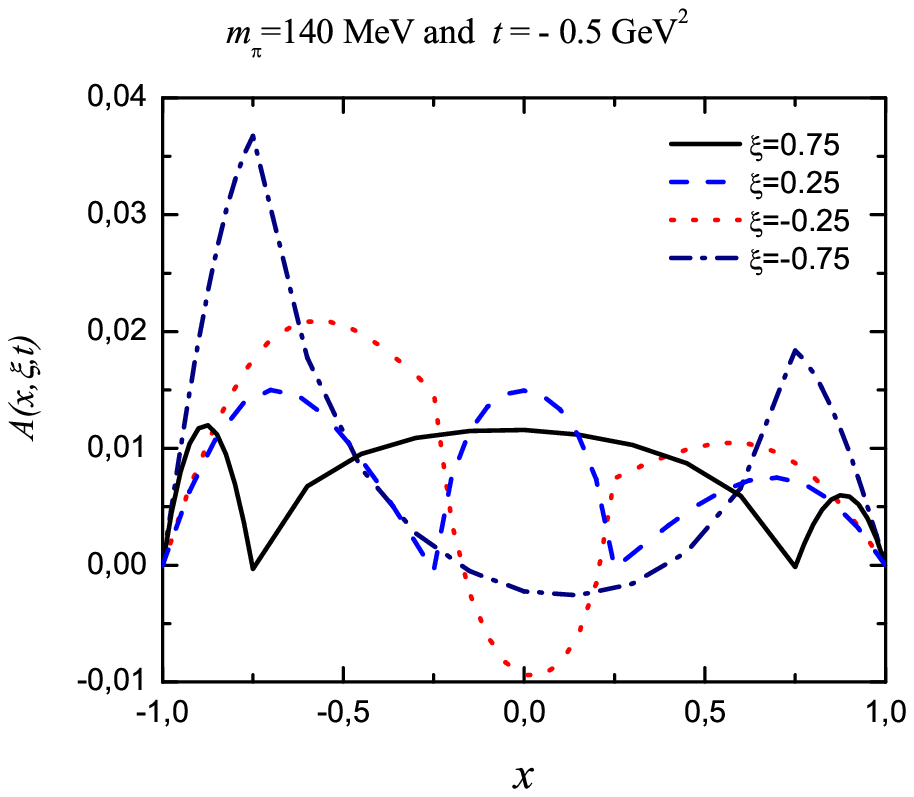}\caption{ The axial TDA in the case
$m_{\pi}=140$ MeV and, respectively, $t=0$ and $t=-0.5$ GeV$^{2}$.}%
\label{axialTDAmpi140}%
\end{figure}

In Figs. \ref{TDAmpi140} and \ref{axialTDAmpi140}, the vector and axial TDAs
are plotted in function of $x$ for several values of $\xi$ and $t$. In these
figures, the $u$-quark contributes to the TDAs in the region of $x$ going from
$-\left\vert \xi\right\vert $ to $1$ and the $d$-antiquark going from $-1$ to
$\left\vert \xi\right\vert .$ Therefore, for $x\in\left[  \left\vert
\xi\right\vert ,1\right]  $ ($x\in\left[  -1,-\left\vert \xi\right\vert
\right]  $) only $u$-valence quarks ($d$-valence antiquarks) are present
(DGLAP regions). Besides, TDAs in the $-\left\vert \xi\right\vert
<x<\left\vert \xi\right\vert $ region (ERBL\ region) receive contributions
from both type of quarks. For the vector TDA, we observe in figure
\ref{TDAmpi140} that the position of the maxima is given by the $\xi$ value,
separating explicitly the ERBL region from the DGLAP regions. The vector TDA
is positive (negative) for negative (positive) values of $x$ with the change
of sign occurring around $x=0$. This change in the sign of the vector TDA is
originated in the presence of the electric charge of the quarks in
Eq.(\ref{3.03}).

The process involved in the calculation of the TDAs allows negative values of
the skewness variable. In the chiral limit, the skewness variable goes from
$\xi=-1$ to $\xi=1,$ for any value of $t$. In the chiral limit, the vector TDA
for a negative value of $\xi$ is equal to the vector TDA for $\left\vert
\xi\right\vert $. This can be seen from the polynomiality expansion in this
limit, Eq. (\ref{even}), which has only even powers of $\xi.$ For the physical
value of the pion mass, negative values of $\xi$ are bounded by $t/\left(
2m_{\pi}^{2}-t\right)  <\xi.$ For each allowed value of $\xi$, we found that
the numerical value of $V\left(  x,-\left\vert \xi\right\vert ,t\right)  $ is
close to $V\left(  x,\left\vert \xi\right\vert ,t\right)  ,$ due to the
smallness of the coefficients of the odd powers of $\xi$ in the polynomial
expansion (\ref{oddeven}).

\begin{figure}[tb]
\centering
\includegraphics [height=7cm]{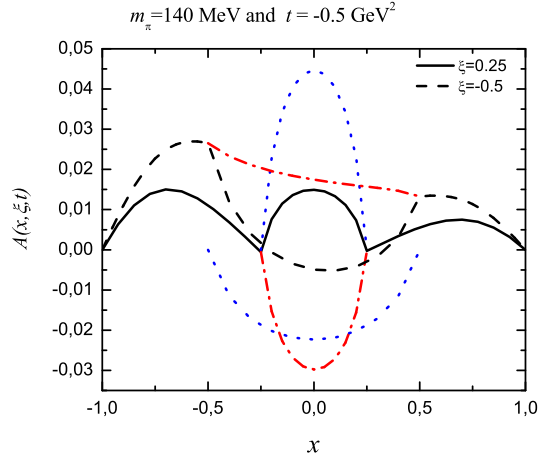}\caption{Contributions to the axial TDA
for both positive ($\xi=0.25,$ plain line) and negative ($\xi=-0.5,$ dashed
line) values of the skewness variable and for $m_{\pi}=140$ MeV and $t=-0.5$
GeV$^{2}$. In each case, and in the ERBL region, the contribution coming from
the first diagram of Fig.~\ref{Fig3} is represented by the dashed-dotted lines
and the non-resonant part of the second diagram of Fig.~\ref{Fig3} is
represented by the dotted lines.}%
\label{comparisonxi}%
\end{figure}

Analyzing the axial TDAs, plotted in Fig.~\ref{axialTDAmpi140}, we observe two
different behaviours depending on the sign of $\xi$. For positive $\xi,$ the
position of the minima is given by the value of the skewness variable while
the position of the maxima is always $x\simeq0$ in the ERBL region and
$x\simeq\pm(1+\xi)/2$ in the DGLAP regions. For negative $\xi,$ the value of
the axial TDA at $x=\pm\xi$ is important and in some cases is a maximum and,
in the ERBL region, $A\left(  x,\xi,t\right)  $ presents a minimum near $x=0.$
As we have previously shown, the axial TDA in the ERBL region receives
contributions from two different diagrams, depicted in Fig.~\ref{Fig3}. In the
second of these diagrams, a virtual quark-antiquark interacting pair in the
pion channel appears. The pion pole, contained in this diagram, has been
subtracted but the remaining non-resonant part contributes to the axial TDA.
We observe that the latter contribution is the dominant one and produces the
maxima around $x=0$ for positives $\xi$ (Fig.~\ref{comparisonxi}). Now, the
axial TDA does not change the sign when we go from negative to positive values
of $x$. In the axial TDA the change of sign originated in the presence of the
electric charge of the quarks in equation (\ref{3.10}) is compensated by the
change of sign between quark and antiquark contributions generated by the
$\gamma_{5}$ operator present in the axial current.

By comparing the plots for different values of the momentum transfer, it is
observed that the amplitudes are lower for higher $(-t)$ values, as it can be
inferred from the decreasing of the form factors with $\left(  -t\right)  $,
connected to the TDAs through the sum rules. By increasing the $(-t)$ value,
not only the width and the curvature of the TDAs are changed but we also
observe that higher values of $\xi$ are preferred, i.e. the sign of the
derivative of the collection of maxima changes passing from a zero momentum
transfer to a non-zero one.

Isospin relates the value of the vector and axial TDAs in the DGLAP regions,%
\begin{equation}
V\left(  x,\xi,t\right)  =-\frac{1}{2}V\left(  -x,\xi,t\right)  ,~~~~A\left(
x,\xi,t\right)  =\frac{1}{2}A\left(  -x,\xi,t\right)  , ~~~~~~~~~~~\left\vert
\xi\right\vert <x<1~~, \label{5.01}%
\end{equation}
being the factor $1/2$ the ratio between the charge of the $u$ and $d$ quarks.
We observe in Figs.~\ref{TDAmpi140} and \ref{axialTDAmpi140} that our TDAs
satisfy these relations. It must be realize that the relation (\ref{5.01})
cannot be changed by evolution.

Regarding the chiral limit, we observe that both the vector and
axial TDAs do not significantly change going from a non-zero pion mass to the
physical mass, except for the change in the lower bound of $\xi$.

Previous studies of pion-photon TDAs have already been done
\cite{Tiburzi:2005nj, Broniowski:2007fs}. In both works, double distributions
have been used. Even if our model is different, a comparison of the results is
still possible. The aim of the author of the first paper \cite{Tiburzi:2005nj}
is to provide some estimates of the vector and axial TDAs on the basis of the
positivity bounds. The order of magnitude of the obtained amplitudes are
similar to ours, but the former are constrained by the $F_{V}$ and $F_{A}$
form factors, through the sum rules. The sum rules are an input imposed in
Ref. \cite{Tiburzi:2005nj} while it is a result in our calculation. The vector
TDA obtained in this paper has some similarities with ours. Nevertheless, it
does not satisfy the isospin relation (\ref{5.01}) due to the different choice
of the $u$-quark and $d$-antiquark distributions, the first one related to the
pion and the second to the photon, used in the saturation of the positivity
bounds. We have also studied the positivity bounds for GPDs and noticed that
it is actually an upper bound that is sometimes very higher than the value of
the GPD itself. The vector TDA obtained in \cite{Tiburzi:2005nj} is rather
peaked at $x=\pm\xi$, whereas we have a smoother behaviour.

A detailed comparison with the results of Ref. \cite{Broniowski:2007fs} is not
easy, due to the choice of the asymmetric notation. Nevertheless the vector
TDA seems to be consistent with our results and some limits, in particular
when $m_{\pi}=0$ and $t=0,$ can be recovered going from one notation to the other.

Regarding our axial TDA, it differs from the two previous calculations
\cite{Tiburzi:2005nj, Broniowski:2007fs} due to the effect of the non-resonant
part of the second diagram of Fig. \ref{Fig3}. This contribution,
corresponding to the last term of Eq.~(\ref{3.11}), is proportional to
$\left(  t-m_{\pi}^{2}\right)  ^{-1}$ but with zero value for the residue. The
presence of this term is crucial in order to obtain the axial form factor
using the sum rule. In fact, its contribution in Eq. (\ref{faalltwists}) can
be easily recognized. Furthermore this term is dominant in the ERBL region as
we can infer from Fig. \ref{comparisonxi}.

\section{Conclusions.}

\label{SecVIC}

In this paper we have defined the pion-photon vector and axial Transition
Distribution Amplitudes using the Bethe-Salpeter amplitude for the pion. In
order to make numerical predictions we have used the Nambu-Jona Lasinio model.
The Pauli-Villars regularization procedure is applied in order to preserve
gauge invariance.

We know from PCAC that the axial current couples to the pion. Therefore, in
order to properly define the axial TDA (all the structure of the incoming
hadron being included in $A\left(  x,\xi,t\right)  $), we need to extract the
pion pole contribution. In so doing, we found that the axial TDA had two
different contributions, the first one related to a direct coupling of the
axial current to a quark of the incoming pion and to a quark coupled to the
outcoming photon and a second related to the non-resonant part of a
quark-antiquark pair coupled with the quantum numbers of the pion.

The use of a fully covariant and gauge invariant approach guaranties that we
will recover all fundamental properties of the TDAs. In this way, we have the
right support, $x\in\left[  -1,1\right]  $, and the sum rules and the
polynomiality expansions are recovered. We want to stress that these three
properties are not inputs, but results in our calculation. The value we found
for the vector form factor in the NJL model is in agreement with the
experimental result \cite{Yao:2006px} whereas the value found for the axial
form factor is two times larger than in \cite{Yao:2006px}. This discrepancy is
a common feature of quark models \cite{Broniowski:2007fs}. Also the neutral
pion vector form factor $F_{\pi\gamma^{\ast}\gamma}(t)$ is well described.
These results allow us to assume that the NJL model gives a reasonable
description of the physics of those processes at this energy regime.

Turning our attention to the polynomial expansion of the TDAs, we have seen
that, in the chiral limit, only the coefficients of even powers in $\xi$ were
non-null for the vector TDA. No constrain is obtained for the axial TDA.
Nevertheless, the NJL model provides simple expressions for the coefficients
of the polynomial expansions in the chiral limit, Eqs. (\ref{4.03}) and
(\ref{coefaxial}).

We have obtained quite different shapes for the vector and axial TDAs. This is
in part, at least for the DGLAP regions, imposed by the isopin relation
(\ref{5.01}). We have pointed out the importance of the non-resonant part of
the qq interacting pair diagram for the axial TDA in the ERBL region.

It is interesting to inquire about the domain of validity of the relations
(\ref{5.01}). These relations are obtained from the isospin trace calculation
involved in the central diagram of Fig. 3. Due to the simplicity of the
isospin wave function of the pion these results are more general than the NJL
model and could be considered as a result of the diagrams under consideration.
These diagrams are the simplest contribution of handbag type.

The Transition Distribution Amplitudes proposed by the authors of
\cite{Pire:2004ie} open the possibility of enlarging the present knowledge of
hadron structure for they generalize the concept of GPDs for non-diagonal
transitions. Calculated here, as a first step, for pion-to-photon transitions,
these new observables should lead to interesting estimates of cross section
for exclusive meson pair production in $\gamma^{\ast}\gamma$ scattering
\cite{Lansberg:2006fv}.

\begin{acknowledgments}
We are thankful to J. P. Lansberg, B. Pire, L. Szymanowsky and V. Vento for
useful discussions. This work was supported by the sixth framework program of
the European Commission under contract 506078 (I3 Hadron Physics), MEC (Spain)
under contracts BFM2001-3563-C02-01, FPA2004-05616-C02-01 and grant
AP2005-5331, and Generalitat Valenciana under contract GRUPOS03/094.
\end{acknowledgments}

\appendix{}

\section{Appendix.}

\label{AppendixA}

In Section \ref{SecIITDA} we have introduced the minimal kinematics
definitions. The explicit expressions for the pion and photon momenta in terms
of the light-front vectors are:%
\begin{align}
p^{\mu}  &  =\left(  1+\xi\right)  \,\bar{p}^{\mu}-\frac{1}{2}\,\Delta
^{\bot\mu}+\frac{1}{2}\left[  P^{2}\,\left(  1-\xi\right)  +\frac{1}%
{2}\,m_{\pi}^{2}\right]  \,n^{\mu}~~,\label{plc}\\
p^{\prime\mu}  &  =\left(  1-\xi\right)  \,\bar{p}^{\mu}+\frac{1}{2}%
\,\Delta^{\bot\mu}+\frac{1}{2}\left[  P^{2}\,\left(  1+\xi\right)  -\frac
{1}{2}\,m_{\pi}^{2}\right]  \,n^{\mu}~~. \label{pplc}%
\end{align}
Here we have $P^{2}=\left[  (p+p^{\prime})/2\right]  ^{2}=m_{\pi}^{2}/2-t/4,$
$\Delta^{\mu}=\left(  p^{\prime}-p\right)  ^{\mu},$ $\Delta^{\bot\mu}=\left(
0,\Delta^{1},\Delta^{2},0\right)  =\left(  0,\vec{\Delta}^{\bot},0\right)  $
and $\Delta^{2}=t.$ The polarization vector of the real photon, $\varepsilon,$
must satisfy the transverse condition, $\varepsilon.p^{\prime}=0,$ and an
additional gauge fixing condition. When deriving Eq. (\ref{2.05}), we need
$\varepsilon.n=\varepsilon^{+}/P^{+}$ to kinematically become higher-twist,
i.e. $\varepsilon.n\rightarrow0$ when $P^{+}\rightarrow\infty$.
The standard gauge fixing conditions, $\varepsilon^{0}=0$ or $\varepsilon
^{+}=0,$ satisfy the previous requirement. In fact, all what we need is that
the components of the polarization vector remain finite when $P^{+}$ goes to infinity.

In Section \ref{SecIIIFTTAD} we use the NJL model. We follow the notation of
Ref. \cite{Theussl:2002xp} and we refer the reader to this paper for more
details. The NJL model considers the lagrangian%
\begin{equation}
\mathcal{L}=\bar{q}\left(  x\right)  \left(  i\rlap{$/$}\hspace*{-0.05cm}%
\partial-\mu_{0}\right)  q\left(  x\right)  +g\left[  \left(  \bar{q}q\right)
^{2}+\left(  \bar{q}i\gamma_{5}\vec{\tau}q\right)  ^{2}\right]  \quad,
\label{A.01}%
\end{equation}
where $\mu_{0}$ is the current quark mass. As it is well known, the first
consequence of the scalar interacting term is to provide a constituent quark
mass, $m,$ different from the current mass. Due to the point-like character of
the interaction, this lagrangian is not renormalizable. We shall use the
Pauli-Villars regularization in order to render the occurring integrals
finite. This means that for integrals like the ones defined in Eq.~(\ref{3.05}%
), we make the replacement
\begin{equation}
\int\frac{d^{4}p}{(2\pi)^{4}}f(p;m^{2})\longrightarrow\int\frac{d^{4}p}%
{(2\pi)^{4}}\sum_{j=0}^{2}c_{j}f(p;m_{j}^{2})\quad, \label{A.02}%
\end{equation}
with $m_{j}^{2}=m^{2}+j\Lambda^{2}$, $c_{0}=c_{2}=1,$ $c_{1}=-2$. Following
Ref.~\cite{Klevansky:1992qe} the regularization parameters $\Lambda$ and $m$
are determined\ by fitting the pion decay constant and the quark condensate
(in the chiral limit). With the conventional values $\left\langle \bar
{u}u\right\rangle =-(250\,%
\operatorname{MeV}%
)^{3}$ and $f_{\pi}=93$ MeV, we get $m=241$ MeV and $\Lambda=859$ MeV.

The pion-quark coupling constant is given by%

\begin{equation}
g_{\pi qq}^{2}=\frac{-1}{12\left(  I_{2}\left(  m_{\pi}^{2}\right)  +m_{\pi
}^{2}\left(  \partial I_{2}\left(  p\right)  /\partial p^{2}\right)
_{p^{2}=m_{\pi}^{2}}\right)  }\quad, \label{A.03}%
\end{equation}
with
\begin{align}
I_{2}\left(  p^{2}\right)   &  =i\int\frac{d^{4}k}{\left(  2\pi\right)  ^{4}%
}\frac{1}{\left(  k^{2}-m^{2}+i\epsilon\right)  \left(  \left(  k+p\right)
^{2}-m^{2}+i\epsilon\right)  }\nonumber\\
&  =\frac{1}{16\pi^{2}}\sum_{j=0}^{2}c_{j}\left\{  \log\frac{m_{j}^{2}}{m^{2}%
}+2\sqrt{\frac{4m_{j}^{2}}{p^{2}}-1}\arctan\frac{1}{\sqrt{\frac{4m_{j}^{2}%
}{p^{2}}-1}}\right\}  . \label{A.04}%
\end{align}
The numerical value of the pion quark coupling constant is $g_{\pi qq}%
^{2}=6.36$ for the physical value of the mass of the $\pi^{+}$ pion, and
$g_{\pi qq}^{2}=6.71$ in the chiral limit.

In the NJL model, the vector form factor is
\begin{equation}
F_{V}^{\pi^{+}}(t)=\frac{8\,N_{c}}{3}\,m\,m_{\pi}\,\frac{g_{\pi qq}}{\sqrt{2}%
}\,I_{3}(p,p^{\prime})\qquad. \label{fv}%
\end{equation}
In order to calculate the form factors we need the expression for the
three-propagator integral. In the particular case where $p^{2}=m_{\pi}^{2}$
and $p^{\prime}{}^{2}=0$, the expression for $I_{3}(p,p^{\prime})$ is
\begin{align}
I_{3}(p,p^{\prime})  &  =i\,\int\,\,\frac{d^{4}k}{(2\pi)^{4}}\,\,\frac
{1}{\left(  k^{2}-m^{2}+i\epsilon\right)  \left(  (p+k)^{2}-m^{2}%
+i\epsilon\right)  \left(  (p^{\prime}+k)^{2}-m^{2}+i\epsilon\right)  }%
\qquad,\nonumber\\
&  =\frac{1}{(4\pi)^{2}}\,\sum_{j=0}^{2}c_{j}\,\int_{0}^{1}dx\,\frac{x}%
{\sqrt{\rho}}\,\mbox{log}\,\frac{x^{2}t+x(1-x)m_{\pi}^{2}-2m_{j}^{2}%
-\sqrt{\rho}}{x^{2}t+x(1-x)m_{\pi}^{2}-2m_{j}^{2}+\sqrt{\rho}}\qquad,
\label{i3}%
\end{align}
with $\rho=t^{2}x^{4}+x^{2}(1-x)^{2}m_{\pi}^{4}+2\,t\,x^{2}(m_{\pi}%
^{2}x(1-x)-2m_{j}^{2}),~t=\left(  p^{\prime}-p\right)  ^{2}$.

The axial form factor involves the three-propagator integral as well, but also
the two-propagator integral given by Eq.~\ref{A.04}
\begin{equation}
F_{A}^{\pi^{+}}(t)=4N_{c}\,m\,m_{\pi}\,g_{\pi qq}\sqrt{2}\,\left(
I_{3}(p,p^{\prime})+\frac{2}{m_{\pi}^{2}-t}\left[  I_{2}(m_{\pi}^{2}%
)-I_{2}(t)\right]  \right)  \qquad, \label{faalltwists}%
\end{equation}

The $\pi^{0}\rightarrow\gamma^{\ast}\gamma$ form factor can be obtained from
the vector form factor through a isospin rotation: $F_{\pi\gamma^{\ast}\gamma
}\left(  t\right)  =\sqrt{2}F_{V}^{\pi^{+}}\left(  t\right)  /m_{\pi}.$ In the
chiral limit and for $t=0$ all the considered form factors have the simple
expression:%
\begin{align}
\left.  \frac{F_{\pi\gamma^{\ast}\gamma}\left(  0\right)  }{\sqrt{2}%
}\right\vert _{m_{\pi}\rightarrow0}  &  =\left.  \frac{F_{V}^{\pi^{+}}\left(
0\right)  }{m_{\pi}}\right\vert _{m_{\pi}\rightarrow0}=\left.  \frac
{F_{A}^{\pi^{+}}\left(  0\right)  }{m_{\pi}}\right\vert _{m_{\pi}\rightarrow
0}=\frac{1}{\sqrt{2}4\pi^{2}f_{\pi}}\left[  1-\frac{2m^{2}}{m^{2}+\Lambda^{2}%
}+\frac{m^{2}}{m^{2}+2\Lambda^{2}}\right] \nonumber\\
&  =0.192\times\left(  1-0.108\right)
\operatorname{GeV}%
^{-1}=0.171%
\operatorname{GeV}%
^{-1}%
\end{align}
The first coefficient of the right hand side is what is expected from the
axial anomaly contribution to $\pi\rightarrow\gamma\gamma$ decay. The term
between brackets has a small correction to the expected value of 1 due to the
finiteness of the regularization masses. In the NJL model not only the quarks,
but also the counter-terms run in the triangle diagram of the axial anomaly.In
a proper renormalizable theory this correction disappears in the limit
$\Lambda\rightarrow\infty.$

In the light-front calculations we need two kind of integrals. The first one is%

\begin{align}
\tilde{I}_{2}\left(  x,\xi,t\right)   &  =i\int\frac{d^{4}k}{(2\pi)^{4}%
}\,\,\frac{\delta\left(  x-1+\frac{k^{+}}{P^{+}}\right)  }{[(k-p_{1}%
)^{2}-m^{2}+i\epsilon][(k-p_{2})^{2}-m^{2}+i\epsilon]}\quad,\nonumber\\
&  =\theta(b-x)\theta(x-a)\,\frac{1}{(4\pi)^{2}}\,\frac{1}{a-b}\,\sum
_{j=1}^{2}c_{j}\,\mbox{log}\,\frac{m^{2}(b-a)^{2}-(b-x)(x-a)(p_{2}-p_{1})^{2}%
}{m_{j}^{2}(b-a)^{2}-(b-x)(x-a)(p_{2}-p_{1})^{2}}\qquad, \label{A.05}%
\end{align}
with $b=$max$\left(  1-p_{1}^{+}/P^{+},1-p_{2}^{+}/P^{+}\right)  $ and
$a=$min$\left(  1-p_{1}^{+}/P^{+},1-p_{2}^{+}/P^{+}\right)  $. The second
light-front integral needed in our calculations is the one defined by Eq.
(\ref{3.05}). This integral can be performed by standard methods, obtaining%
\begin{equation}
\tilde{I}_{3v}\left(  x,\xi,t\right)  =\frac{1}{32\,\pi^{2}}\,\sum_{j=0}%
^{2}c_{j}\,\frac{1}{\sqrt{D}}\,\mbox{log}\,\frac{-\frac{t}{2}(1-x)+C+\sqrt{D}%
}{-\frac{t}{2}(1-x)+C-\sqrt{D}}\qquad,
\end{equation}
with
\begin{align}
D  &  =\left(  \frac{m_{\pi}^{2}}{2}(\xi-x)-\frac{t}{2}(1-x)\right)
^{2}+m_{j}^{2}(1-\xi)\left(  2m_{\pi}^{2}\xi-t(1+\xi)\right)  \qquad
,\nonumber\\
C  &  =\left\{
\begin{array}
[c]{ccl}%
\frac{m_{\pi}^{2}}{2}(\xi-x)+m_{j}^{2}\frac{(1-\xi^{2})}{(1-x)}\quad, &  &
|\xi|<x<1,\nonumber\\
\nonumber &  & \\
-\frac{m_{\pi}^{2}}{2}(\xi-x)+m_{j}^{2}\frac{2\xi(1+\xi)}{(x+\xi)}\quad, &  &
-|\xi|<x<|\xi|~~~~~~~\xi>0\\
\nonumber &  & \\
-\frac{m_{\pi}^{2}}{2}(\xi+x)-m_{j}^{2}\frac{2\xi(1-\xi)}{(x-\xi)}\quad, &  &
-|\xi|<x<|\xi|~~~~~~~\xi<0\quad.
\end{array}
\right.
\end{align}

As an aplication of the expression (\ref{A.05}) we can evaluate the PDA. From
its definition in Eq. (\ref{2.08}) we have%
\begin{equation}
\phi\left(  x\right)  =-\frac{4\,N_{c}\,m\,g_{\pi qq}}{f_{\pi}}i\int
\frac{d^{4}k}{(2\pi)^{4}}\,\,\frac{\delta\left(  x-1+\frac{k^{+}}{p^{+}%
}\right)  }{[k^{2}-m^{2}+i\epsilon][(k-p)^{2}-m^{2}+i\epsilon]}~~.
\end{equation}
This integral is of the type of (\ref{A.05}) with $a=0$ and $b=1.$ Therefore%
\begin{equation}
\phi\left(  x\right)  =\frac{N_{c}\,m\,g_{\pi qq}}{4\,\pi^{2}\,f_{\pi}%
}\,\theta\left(  x\right)  \,\theta\left(  1-x\right)  \sum_{j=1}^{2}%
c_{j}\,\log\,\frac{m^{2}-m_{\pi}^{2}\,(1-x)\,x}{m_{j}^{2}-m_{\pi}%
^{2}\,(1-x)\,x}~~.
\end{equation}
As it is expected for the PDA, $x$ runs from 0 to 1. We can test our result
verifying that in the chiral limit $\phi\left(  x\right)  =\theta\left(
x\right)  \,\theta\left(  1-x\right)  ,$ as it is well known for the NJL
model. This peculiarity of the NJL model in this limit is present also in the
quark valence distribution, which becomes as simple as $q(x)=\theta\left(
x\right)  ~\theta(1-x)$. In Ref. \cite{Noguera:2005cc} is discussed how this
is, in fact, a quite reasonable approximation of realistic models.

\end{document}